\documentclass[11pt,a4paper]{article}

\usepackage[utf8]{inputenc}

\usepackage{url}
\usepackage[final,dvips]{graphicx}
\usepackage{epsfig}
\usepackage{wrapfig}
\usepackage{float}
\usepackage[hang,small,bf]{caption}
\newcounter{print}
\newcounter{commentbox}
\usepackage[dvips]{graphicx}

\usepackage[scaled=.92]{helvet} 
\usepackage{courier}            

\usepackage{color}

\usepackage{amsmath, amsthm, amssymb}

\def\lsim{\mathrel{\rlap{\lower1.2pt\hbox{\hskip0.6pt${\scriptstyle\sim}$}}\raise2pt\hbox{${\scriptstyle<}$}}}

\newcommand{\bm}[1]{\mbox{\boldmath$#1$}}

\begin{document}


\title{PCAC and coherent pion production by low energy neutrinos}

\author{Ch. Berger
\thanks{\noindent email: berger@rwth-aachen.de}
\\{\small I. Physikalisches Institut der RWTH Aachen University, Germany} \and
L. M. Sehgal
\thanks{email: sehgal@physik.rwth-aachen.de}
\\{\small Institut f{\"u}r Theoretische Physik (E) der  RWTH Aachen University, Germany}
\date{}}
\maketitle


\begin{abstract}
Coherent $\pi^+$ and $\pi^\circ$ production in low energy neutrino reactions is discussed in the
framework of the partially conserved axial vector current theory (PCAC). The role of lepton mass effects in suppressing the $\pi^+$ production is emphasized. Instead of using models of
pion nucleus scattering,  the available data on pion Carbon scattering are implemented for
an analysis of the PCAC prediction. Our results agree well with the published upper limits
for $\pi^+$ production but are much below the recent MiniBooNE result for $\pi^\circ$
production.

\flushright{PACS 13.15.+g 11.40.Ha 25.80.Dj }
\end{abstract}


\section{Introduction}
\label{intro}
The availability of high intensity neutrino beams with energies up to a few GeV
opens the way to precise investigation of neutrino oscillations. Essential for these experiments is a detailed understanding of all low energy neutrino reactions especially single pion production in charged (CC) and neutral (NC) current reactions. Coherent pion production off nuclei, e.g. in 
$\nu_\mu +^{12}{\rm C}\rightarrow \nu_\mu +^{12}{\rm C}+\pi^\circ$
 constitutes an especially interesting subsample not only because it is a significant background
to the $\nu_\mu\to\nu_e$ oscillation search but also because it is deeply rooted in fundamental physics via Adler's PCAC theorem~\cite{Adler1,Adler2}
which connects forward neutrino scattering with the pion nucleon cross section.

\section{PCAC and forward lepton theorem}
\label{formalism}
Our starting point is the general
expression for inelastic neutrino scattering\footnote{$G_F$ and $\theta_C$ are the Fermi coupling constant
and the Cabbibo angle.}
\begin{equation}
  \frac{d\sigma^{CC}}{dQ^2 dy} = \frac{G_F^2\cos^2\theta_C}{4 \pi^2 } \kappa E
 \frac{Q^2}{|\bm{q}|^2} \left[ u^2 \sigma_L + v^2 \sigma_R + 2uv \sigma_S \right]
  \label{eq:1}
\end{equation}
already derived by Lee and Yang~\cite{LeeYang} in 1962 for zero mass of the outgoing lepton.
The momentum and energy transfer between incoming neutrino and outgoing lepton is given by $\bm{q}$ and $\nu=E-E'$ with
$y=\nu/E$. As usual $Q^2=-q^2$ denotes the four-momentum transfer squared, $Q^2=\bm{q}^2-\nu^2$ while
$\kappa=(W^2-M_N^2)/2M_N$ for a hadronic system with invariant mass $W$ emerging from a nucleus or
nucleon with mass $M_N$. The kinematical factors $u,v$ are given by
$u,v=(E+E'\pm |\bm{q}|)/2E$.

The cross sections $\sigma_{L,R,S}$ (where $L,R,S$ stands
for \emph{Left, Right, Scalar}) are unknown functions of $Q^2$ and $W$
which have to be measured or calculated in theoretical models. It is well known that for $Q^2\to 0$ only the term with
$\sigma_S$ in (\ref{eq:1}) survives because the overall factor $Q^2$ is compensated by a factor
$1/Q^2$ in the scalar cross section. In this limit  PCAC predicts~\cite{Ravndal}
\begin{equation}
 \sigma_S=\frac{|\bm{q}|}{\kappa Q^2}f_\pi^2\sigma_{\pi N}\enspace .
\label{eq:2}\end{equation}
Here $f_\pi$ is the pion decay constant (130.7 MeV) and $\sigma_{\pi N}(W)$ the  pion
nucleon (or nucleus) cross section for the hadronic final state under consideration.

The resulting formula for forward inelastic neutrino scattering is\footnote{$\theta_l$ denotes
the laboratory angle of the outgoing lepton.}
\begin{equation}
\frac{d\sigma^{CC}}{dQ^2 dy}\bigg|_{\theta_l\to 0} = \frac{G_F^2  \cos^2 \theta_C f_\pi^2}{2\pi^2 }\frac{E}{|\bm{q}|}
uv\sigma_{\pi N}(W) \enspace .
\label{eq:3}\end{equation}
In the presence of lepton mass $m_l$ there is a correction to this formula caused by the pion pole term
in the hadronic axial vector current. Including, in addition, an axial vector form factor
$G_A(Q^2)$ to describe the variation of the cross section at small values of
$Q^2$ around the forward direction, one obtains the result given by
Kopeliovich and Marage~\cite{Kop}
\begin{equation}
\frac{d\sigma^{CC}}{dQ^2 dy} = \frac{G_F^2  \cos^2 \theta_C f_\pi^2}{2\pi^2 }\frac{E}{|\bm{q}|}
uv\left[\left(G_A-\frac{1}{2}\frac{Q^2_{\rm min}}{Q^2+m_\pi^2}\right)^2+
\frac{y}{4}(Q^2-Q^2_{\rm min})\frac{Q^2_{\rm min}}{(Q^2+m_\pi^2)^2}\right]\sigma_{\pi N}\enspace ,
\label{eq:4}
\end{equation}
where $Q^2_{\rm min}= m_l^2y/(1-y)$ is the high energy approximation
to the true minimal $Q^2$.  The axial vector form factor $G_A$ is defined by
\begin{equation}
 G_A=\frac{m_A^2}{Q^2+m_A^2}\enspace .
\label{eq:5}\end{equation}
with  a typical value for the axial vector meson mass $m_A$ of
0.95 GeV; see however the extensive discussion in the literature~\cite{GAlit}.

The first term inside the rectangular brackets of (\ref{eq:4}) corresponds to outgoing
muons with negative helicity (helicity nonflip) whereas the second term is 
the helicity flip contribution which vanishes  at $0^\circ$ scattering angle.
This helicity structure was also found by Adler~\cite{Adler2} and
Piketty and Stodolsky~\cite{Stod}.
With $G_A=1$ the expression inside
the rectangular brackets represents the \emph{Adler screening factor}
which was invoked in~\cite{Sehgal,BRS} as a possible explanation of the dip in
CC reactions at low $Q^2$, resulting from the destructive interference of the pion pole. 

For neutrino scattering off nuclei $N$ the coherent pion channel 
$\nu_\mu N\to \mu^-\pi^+N$  
has special interest, since the PCAC formula (\ref{eq:4}) predicts the cross section to
be proportional to the elastic cross section $\pi N\to \pi N$. This hadronic process
is strongly enhanced in the forward direction because of the coherent action of the
$A$ nucleons in the nucleus, with an amplitude
 $\sim A$ at $0^\circ$ scattering angle. This effect then implies the enhancement of forward going pions
in the neutrino reaction, which is a characteristic signature of coherent pion
production, distinguishing it from possible incoherent backgrounds.
An early analysis of pion production by neutrinos, that included a discussion of the kinematical region termed coherent, was given in~\cite{Gershtein}.

\section{Application to coherent pion production}

In the following we investigate in more detail coherent single pion production.
Assuming that the derivation given above also holds for the differential cross section
one gets for the CC reaction $\nu_\mu N\to \mu^-\pi^+ N$
\begin{eqnarray}
\frac{d\sigma^{CC}}{dQ^2 dy dt}= \frac{G_F^2  \cos^2 \theta_C f_\pi^2}{2\pi^2 }\frac{E}{|\bm{q}|}
uv\left[\left(G_A-\frac{1}{2}\frac{Q^2_{\rm min}}{Q^2+m_\pi^2}\right)^2+
\frac{y}{4}(Q^2-Q^2_{\rm min})\frac{Q^2_{\rm min}}{(Q^2+m_\pi^2)^2}\right]
\nonumber\\
\times \frac{d\sigma (\pi^+ N\to\pi^+ N)}{dt}
\label{eq:6}\end{eqnarray}
where  $t$ is the modulus of the four-momentum transfer squared between
incoming virtual boson  and outgoing pion.  For calculating the NC reaction $\nu N\to \nu \pi^\circ N$ one has to set 
$m_l=0, \,\,\theta_C=0$
and divide the right hand side of the resulting equation by 2, because $f_{\pi^\circ}=f_\pi/\sqrt{2}$.
We thus obtain
\begin{equation}
 \frac{d\sigma^{NC}}{dQ^2 dy dt}
 = \frac{G_F^2 f_\pi^2}{4\pi^2 }\frac{E}{|\bm{q}|}
uv G_A^2\frac{d\sigma (\pi^\circ N\to\pi^\circ N)}{dt}\enspace .
\label{eq:7}
\end{equation}
For isoscalar targets $d\sigma (\pi^+ N\to\pi^+ N)$ equals
 $d\sigma (\pi^\circ N\to\pi^\circ N)$.

In the widely used Rein-Sehgal (RS) model~\cite{RS} the kinematical factors on the right hand side of (\ref{eq:7}) have been 
 evaluated for $Q^2=0$
resulting in the simpler expression
\begin{equation}
\frac{d\sigma^{NC}}{dQ^2 dy dt} = \frac{G_F^2 f_\pi^2}{4\pi^2}\frac{1-y}{y}
G_A^2\frac{d\sigma(\pi^\circ N\to\pi^\circ N) }{dt}
\label{eq:8}\end{equation}
for the cross section.

We have compared (\ref{eq:7}) to (\ref{eq:8}) by performing a Monte Carlo integration
using the illustrative ansatz 
\begin{equation}
 \frac{d\sigma(\pi N\to\pi N) }{dt}=\sigma_0be^{-bt}
\label{eq:9}\end{equation}
with energy independent coefficients $\sigma_0=80$ mb and $b=45$ GeV$^{-2}$ for elastic $\pi^\circ$ or
$\pi^+$ scattering off
Carbon nuclei. (This ansatz will be motivated in the next section.) 

\begin{figure}
\epsfig{figure=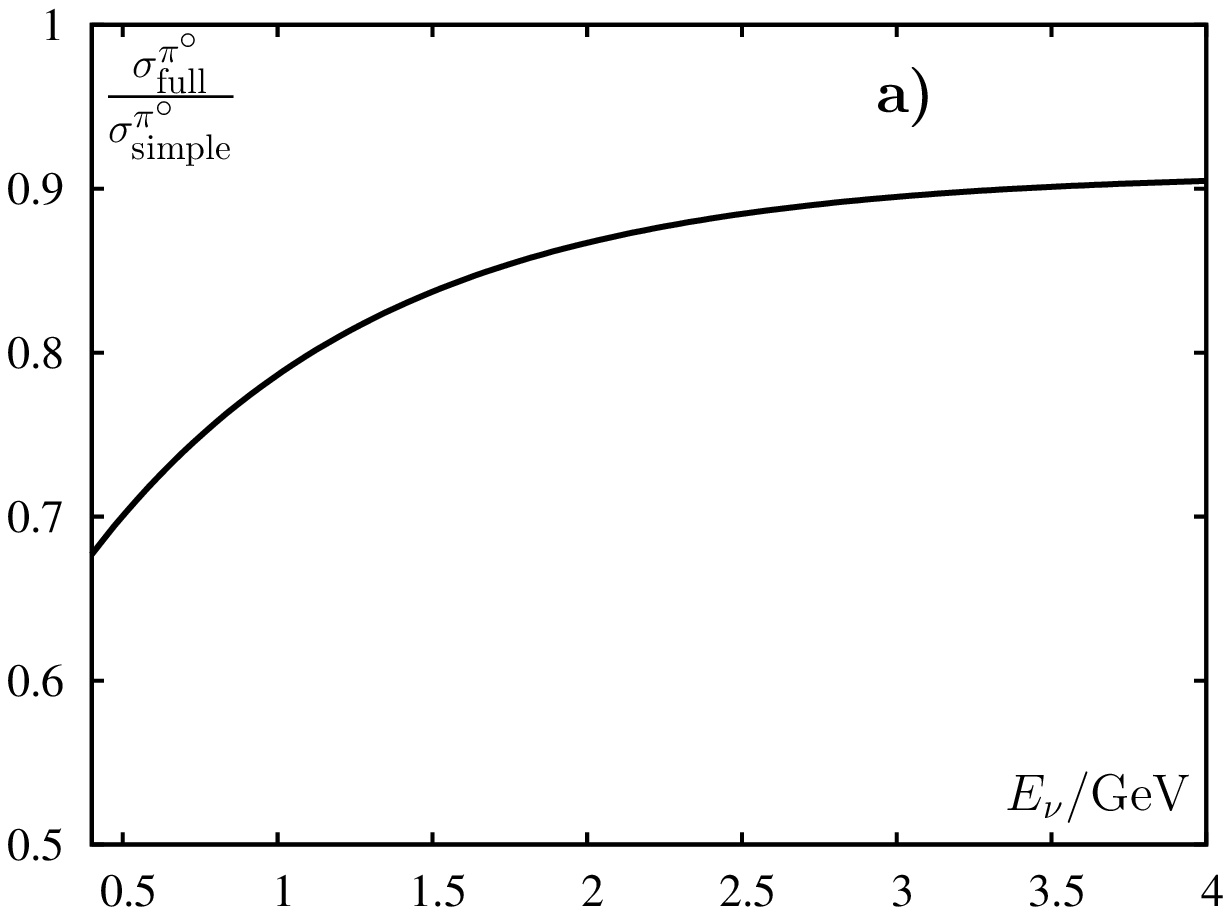,width=0.48\textwidth}\hspace{0.2cm}
\epsfig{figure=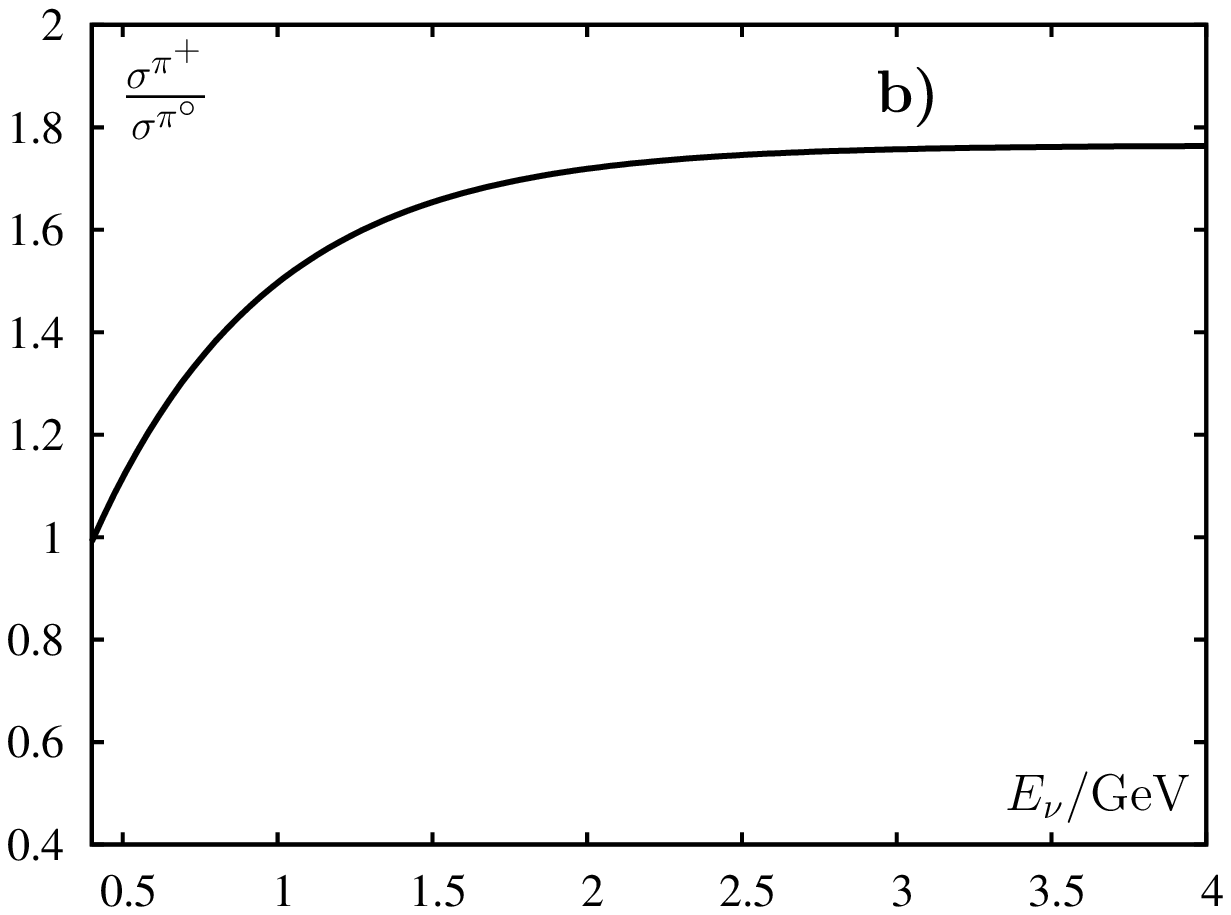,width=0.48\textwidth}
\vspace{0.5cm}

\caption[fig1]{a) Ratio $\sigma^{\pi^0}_{\rm full}/\sigma^{\pi^0}_{\rm simple}$ 
of the integrated cross sections of (\ref{eq:7}) and (\ref{eq:8}) versus 
the energy of the incoming neutrino. b)  Ratio $\sigma^{\pi^+}/\sigma^{\pi^0}$ 
of the integrated cross sections (\ref{eq:6}) and (\ref{eq:7}) versus 
the energy of the incoming neutrino. }\label{fig1}
\end{figure}

The result of the integration is shown in fig.\ref{fig1}a where the ratio 
$\sigma^{\pi^0}_{\rm full}/\sigma^{\pi^0}_{\rm simple}$ is plotted versus the energy of the incoming
neutrino. Here $\sigma^{\pi^0}_{\rm full}$ stands for the integral of (\ref{eq:7}) and
 $\sigma^{\pi^0}_{\rm simple}$ for the integral of (\ref{eq:8}). This figure demonstrates
that at low energies the neutrino cross section is substantially reduced by using the
complete kinematical factors of (\ref{eq:7}).

We also compared $\pi^+$ production to $\pi^\circ$ production by integrating
(\ref{eq:6}) and (\ref{eq:7}) using identical form factors and pion nucleus cross sections.
Apart from the factor $\cos^2 \theta_C$ one would naively expect a ratio of 2 according to the ratio of the pion decay constants.
Fig.\ref{fig1}b shows however a remarkable  violation of this isospin symmetry due
to lepton mass effects contained in (\ref{eq:6}). The dominant cause for the variation
observed in the figure is the reduced phase space of CC reactions. The Adler screening factor
reduces the CC cross section by further 10\% at $E=0.6$ GeV and 4\% at $E=2$ GeV. The influence
of the numerical value of $m_A$ on the cross section ratio is neglegible. Even setting
$G_A=1$ changes the ratio by less then 2\%. 


\section{The elastic pion Carbon cross section}
In \cite{RS} a model  has been presented which calculates elastic pion nucleus scattering
from pion nucleon scattering via
\begin{equation}
 \frac{d\sigma(\pi N\to\pi N) }{dt}=A^2\frac{d\sigma_{\rm el}}{dt}\Big|_{t=0}e^{-bt}F_{\rm abs}\enspace .
\label{eq:10}\end{equation}
Here the pion on the left hand side can be charged or neutral.
The differential elastic pion nucleon cross section in  forward direction on the right hand
side is determined
via the optical theorem (neglecting a possible real part of the scattering amplitude)
\begin{equation}
 \frac{d\sigma_{\rm el}}{dt}\Big|_{t=0}=\frac{1}{16\pi}
\left(\frac{\sigma^{\pi^+p}_{\rm tot}+\sigma^{\pi^-p}_{\rm tot}}{2}\right)^2
\end{equation}
 and the slope $b$ of the exponential $t$-distribution is taken from the optical model relation
\begin{equation}
 b=\frac{1}{3}R_0^2A^{2/3}\enspace
\end{equation}
with e.g. $R_0=1.057$ fm. $F_{\rm abs}$ describes the average attenuation  of a pion emerging
from a sphere of nuclear matter with radius $R_0A^{1/3}$ resulting in
\begin{equation}
 F_{\rm abs}=\exp{\left(-\frac{9A^{1/3}}{16\pi R_0^2}\sigma_{\rm inel}\right)}
\label{eq:13}\end{equation}
with
\begin{equation}
\sigma_{\rm inel}=\frac{\sigma^{\pi^+p}_{\rm inel}+\sigma^{\pi^-p}_{\rm inel}}{2}\enspace .
\end{equation}

As an example we calculate the total elastic pion Carbon cross section via
\begin{equation}
 \sigma_{\rm el}(\pi\, ^{12}C\to\pi\, ^{12}C)=\frac{A^2 F_{\rm abs}}{16\pi b}
\left(\frac{\sigma^{\pi^+p}_{\rm tot}+\sigma^{\pi^-p}_{\rm tot}}{2}\right)^2\enspace .
\label{eq:17}\end{equation}
The total pion nucleon cross sections are available as computer readable files~\cite{PDG}.
The data were fitted by a superposition of Breit Wigner functions and a Regge inspired
term $a_0+a_1/\sqrt{|\bm{p}_\pi|}$ with $|\bm{p}_\pi|$ denoting the pion laboratory momentum.
A similar fit to the elastic cross sections finally yields $\sigma_{\rm inel}=\sigma_{\rm tot}
-\sigma_{\rm el}$ which is used for calculation of $F_{\rm abs}$.

\begin{figure}
\begin{minipage}{0.595\textwidth}

\includegraphics[width=0.9\textwidth]{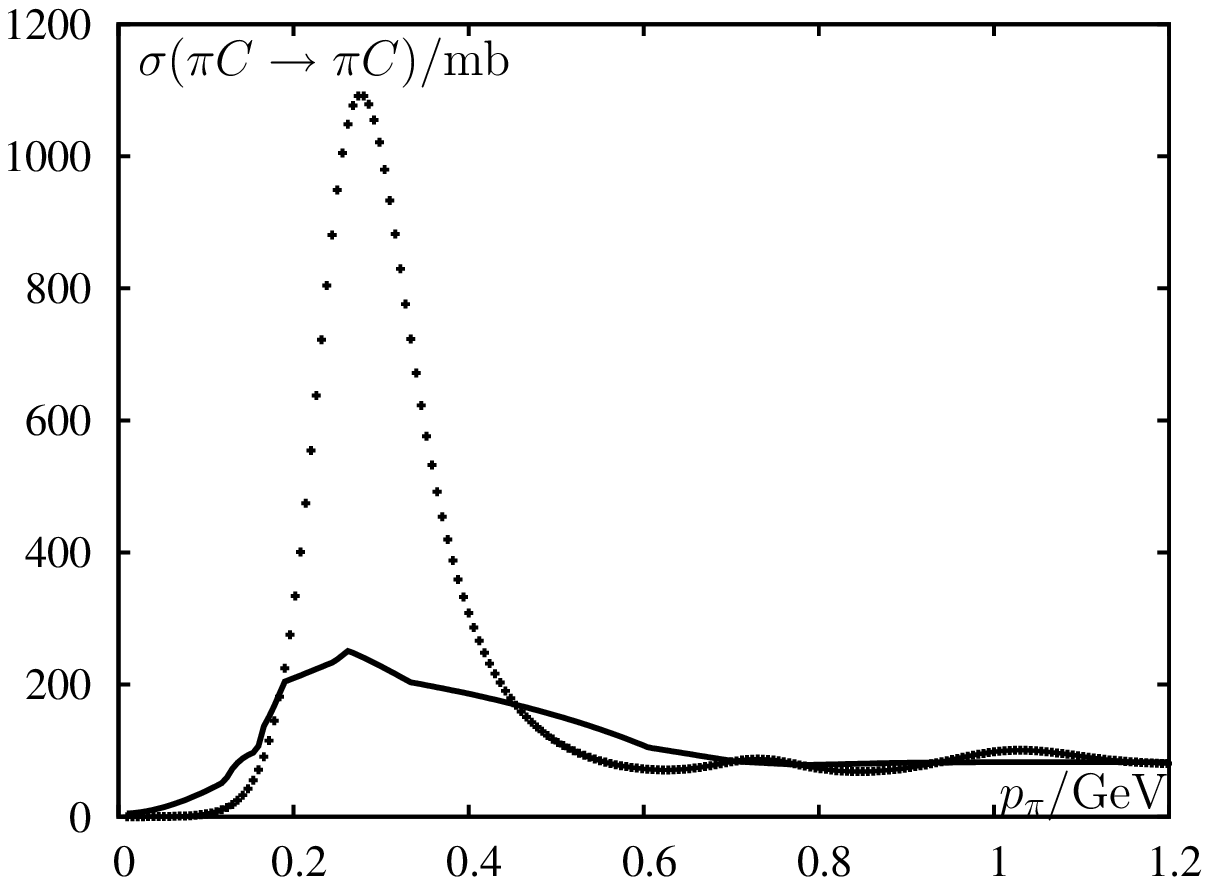}
\vspace{0.5cm}

\end{minipage}
\begin{minipage}{0.4\textwidth}
\vspace{-0.8cm}
\hfill
{\small
 \begin{tabular}{r|r|r}
$T_\pi$&$A_1$&$b_1$\\
GeV&mb/GeV$^2$&1/GeV$^2$\\
\hline
0.076&11600&116.0\\
0.080&14700&109.0\\
0.100&18300&89.8\\
0.148&21300&91.0\\
0.162&22400&89.2\\
0.226&16400&80.8\\
0.486&5730&54.6\\
0.584&4610&55.2\\
0.662&4570&58.4\\
0.776&4930&60.5\\
0.870&5140&62.2\\
\hline
\end{tabular}
}
\end{minipage}
\caption{Total elastic pion Carbon cross section versus
pion laboratory momentum. The dotted line represents the Rein-Sehgal model according to
(\ref{eq:17}), the solid line is derived from pion Carbon data as explained in 
the text. The table on the right hand side contains the coefficients
$A_1,b_1$ of (\ref{eq:16}).}
\label{fig2}
\end{figure}

The dotted line of fig.\ref{fig2} shows the result.
The fact that this cross section is derived from a simple (classical) ansatz expressed by
(\ref{eq:10}) and (\ref{eq:13})
raises doubts about its validity as a description of pion-nucleus scattering in the resonance region.
An alternative approach to coherent pion nucleus interaction based on the Glauber
model was proposed by Bel'kov and Kopeliovich~\cite{Belkow}. Its numerical results were
similar to those in the RS model at least at high energies.
The experimental groups use
detailed Monte Carlo routines to simulate the scattering and absorption of the  pion
inside the nucleus~\cite{Mboone,Sciboone}.

Following the PCAC route  we have tried to circumvent the uncertainties in
modelling nuclear processes by direct appeal to 
data on pion nucleus elastic scattering, see also~\cite{Paschos}.
For Carbon targets this can be
done easily because $\pi^+$ and $\pi^-$ data on differential and total cross sections exist for pion kinetic energies $T_\pi$ from
30 to 870 MeV. They have been subjected to phase shift analyses yielding up to 21
complex phase shifts per energy~\cite{Schlaile}. Neglecting electromagnetic
effects these phase shifts can be used to compute the elastic strong interaction
cross section $d\sigma_{\rm el}/dt$ for pion Carbon scattering in a straightforward manner. 

The phase shifts accurately reproduce even tiny effects like secondary peaks in the angular
distribution. We have checked that except for the lowest two kinetic energies of 30 and 50
MeV it suffices to parametrize the cross section by the simple ansatz
\begin{equation}
\frac{d\sigma_{\rm el}}{dt}=A_1e^{-b_1t}
\label{eq:16}\end{equation}
with energy dependent coefficients $A_1,b_1$, which are listed in
the table on  the right hand side of fig.\ref{fig2}.  For energies between the measured data points these
coefficients are linearly interpolated which is the reason for the 
zig-zag structure of the solid line in fig.\ref{fig2}. It is obvious  that
$\sigma_{\rm el}$ from pion Carbon data is much below the RS model in the resonance
region. At the same time one observes that 
as $|\bm{p}_\pi|$ approaches 1 GeV, the two  curves become very similar with
$\sigma_{\rm el}\approx 80$ mb. This finally justifies the ansatz (\ref{eq:9}). It also suggests that 
the RS hadronic model fails in the region of the $\Delta$ resonance, but may be a valid description
at higher energies.

\section{Results}

\begin{figure}
\epsfig{figure=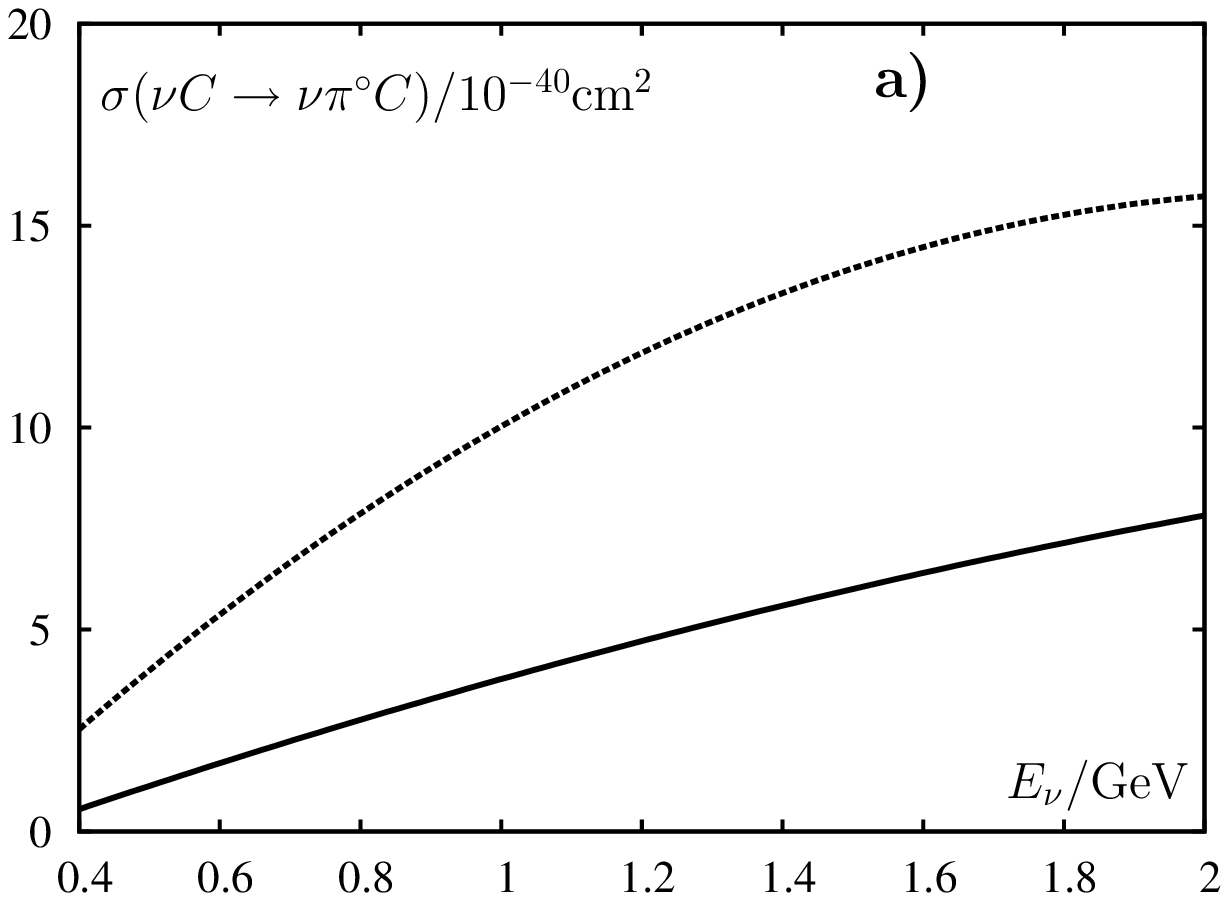,width=0.48\textwidth}\hspace{0.5cm}
\epsfig{figure=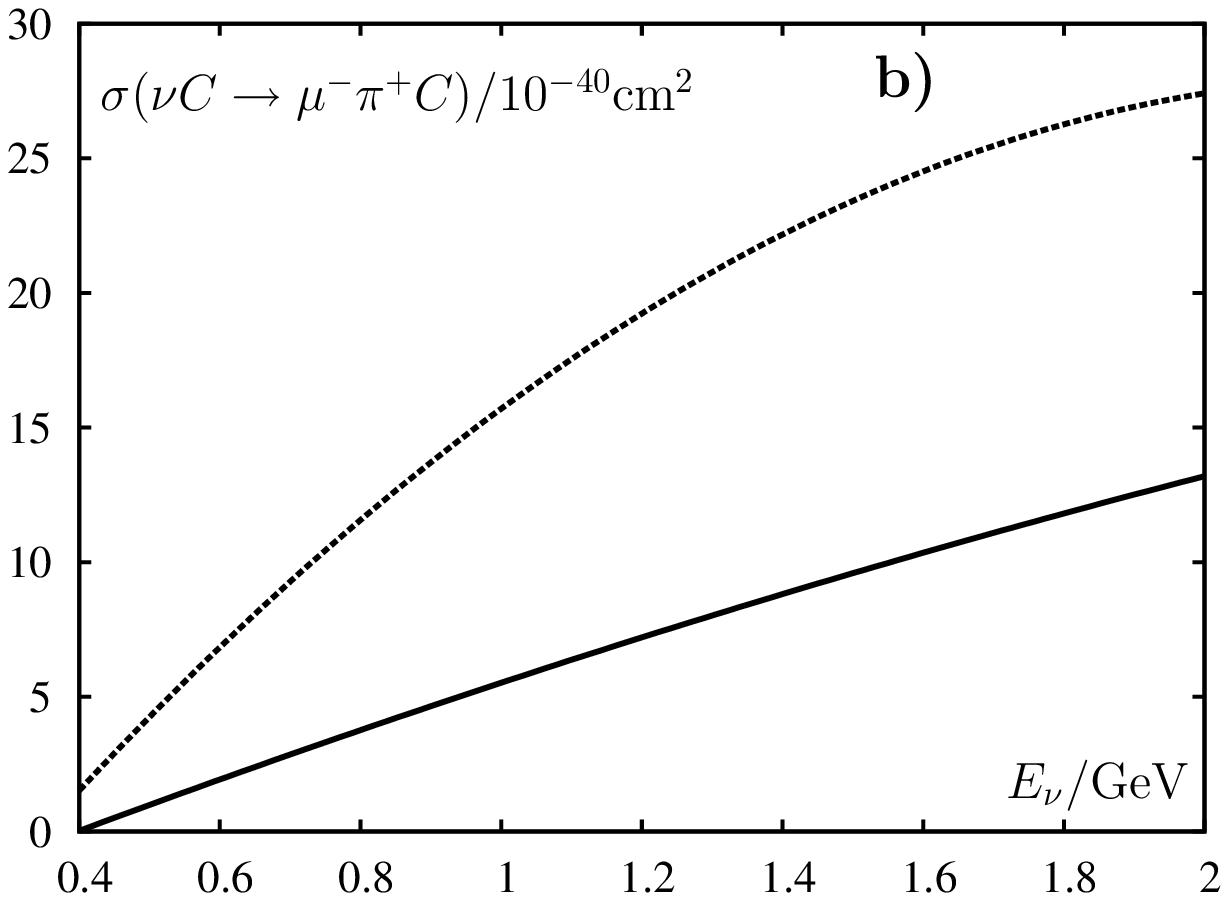,width=0.48\textwidth}
\vspace{0.5cm}
\caption[fig4]{Cross section per nucleus of coherent $\pi$ production
by neutrinos off Carbon nuclei, a) NC reaction $\nu_\mu +^{12}{\rm C}\rightarrow \nu_\mu +^{12}{\rm C}+\pi^\circ$, b) CC reaction $\nu_\mu +^{12}{\rm C}\rightarrow \mu^- +^{12}{\rm C}+\pi^+$.
The results in units of $10^{-40}$ cm$^2$
are plotted versus the neutrino energy in GeV. The upper curve is calculated using
the hadronic RS model, the lower curve using our parametrization
of pion Carbon scattering data.}
\label{fig3}\end{figure}

\begin{figure}
\vspace{-1cm}
\centering\epsfig{figure=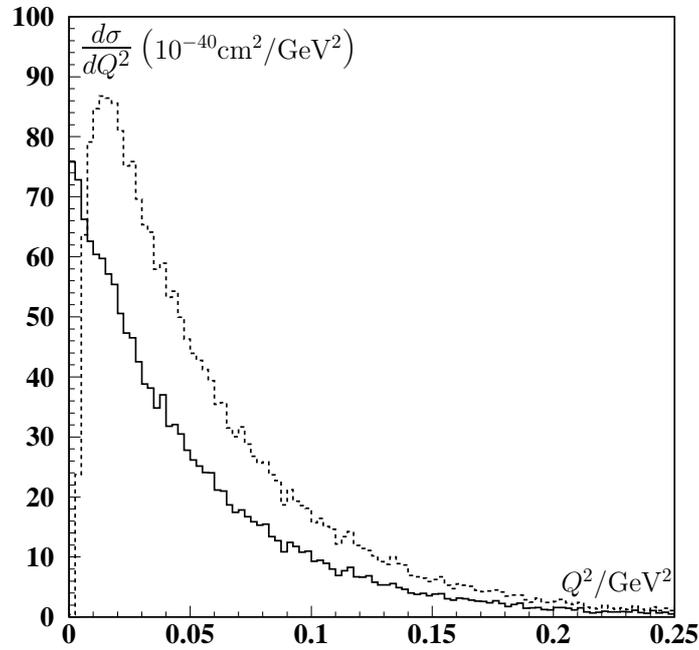,width=0.6\textwidth}
\caption{Differential cross section 
 $d\sigma/dQ^2$ per nucleus for coherent single pion production off Carbon nuclei.
The data are obtained by integrating  (\ref{eq:7}) using Carbon data for
$\sigma_{\pi N}$. The neutrino energy is 1 GeV. Solid line is for the NC reaction, the dashed line
for the CC reaction.}
\label{fig4}\end{figure}

We are now ready to integrate the  cross section (\ref{eq:6}) for
the two different models of pion Carbon scattering discussed in the last section. The results are
plotted versus the neutrino energy in fig.\ref{fig3}a for $\pi^0$ production
and in fig.\ref{fig3}b for $\pi^+$ production.
In obtaining the lower curves the empirical pion Carbon cross sections were calculated by assuming the
coefficients in the last line of the table on the right hand side of fig.\ref{fig2} to be valid up to $T_\pi=1.7$ GeV. An error of $30\%$ in this assumption results in a cross section error of
$6\%$ at $E=2$ GeV. 

The curve using Carbon data is a factor
of 3 to 2  below the
curve obtained by applying the RS hadronic model. 
Cross sections for NC and CC coherent single pion production on Carbon
have also been calculated using an ansatz  
based mainly on the
microscopic process $\nu p\to \mu^- \Delta^{++}$ and its modification
in the nuclear environment~\cite{alvarez,alvarez1,Singh,Amaro}.
(For an early reference to this subject see~\cite{Bell}.)
Remarkably our calculations
agree well with the corresponding results given in~\cite{Singh,Amaro} based on 
a very different approach to coherent neutrino scattering.
The predicted cross sections of~\cite{Paschos} depend sensitively on a cut paarameter $\xi$.
Referring to footnote 41 of~\cite{Paschos} with $\xi=1$ the results
are close to the ones obtained in this paper.
The differential cross sections
$d\sigma/dQ^2$ or $d\sigma/d\cos\theta_l$ are more sensitive to details of the theoretical models.
We give in fig.\ref{fig4}  our prediction for $d\sigma/dQ^2$  at a neutrino energy of 1 GeV. 
For the CC reaction a pronounced dip in forward direction is seen which is mainly due to the
Adler screening factor contained in the rectangular brackets of (\ref{eq:7}).  

Comparing to experimental results we first discuss $\pi^+$ production.
At a neutrino energy of 1.3 GeV the K2K experiment~\cite{K2K} has placed
an upper limit of $0.60\times 10^{-2}$
on the cross section ratio of coherent pion production  $\sigma_{\rm coh}^{\pi^+}$
to the total CC
cross section $\sigma_\nu^{CC}$ at $90\%$ confidence level (CL).
Using the K2K  cut in the muon momentum ($p_\mu > 0.45$ GeV) our
prediction is $\sigma_{\rm coh}^{\pi^+}=0.62\times 10^{-40}$ cm$^2$
per nucleon. With $\sigma_\nu^{CC}=107\times 10^{-40}$cm$^2$
as quoted in~\cite{K2K} we obtain
$\sigma_{\rm coh}^{\pi^+}/\sigma_\nu^{CC}=0.58\times 10^{-2}$ which
is consistent with the K2K result.

The SciBooNE experiment~\cite{Sciboone} 
has measured an upper limit of $0.67\times 10^{-2}$ at 1.1 GeV
($90\%$ CL) 
on the same cross section ratio. Assuming the $\sigma_\nu^{CC}$ value
quoted in~\cite{K2K} to scale linearly with $E$ we
obtain $\sigma_\nu^{CC}/E=82.3\times 10^{-40}$ cm$^2$/GeV 
in the low energy regime and  
$\sigma_{\rm coh}^{\pi^+}/\sigma_\nu^{CC}=0.58\times 10^{-2}$
taking $\sigma_{\rm coh}^{\pi^+}$ from fig.\ref{fig3}b.
At 2.2 GeV the SciBooNE upper limit is $1.36\times 10^{-2}$
to be compared with a calculated ratio of $0.68\times 10^{-2}$.
The latter value requires a slight extrapolation of fig.\ref{fig3}b.

The improved model for coherent pion production is thus in good agreement with
the $\pi^+$ measurements.
It is possible that the lower cross section for coherent $\pi^+$ production also 
reduces the discrepancy  at small
$Q^2$ in CC reactions noted by the MiniBooNE collaboration~\cite{Mboone2}.

Turning to $\pi^\circ$ production the published
coherent fraction $\sigma_{\rm coh}/(\sigma_{\rm coh}+\sigma_{\rm incoh})$
of the MiniBooNE experiment~\cite{Mboone} is $(19.5\pm 1.1 {\rm (stat)}
\pm 2.5 {\rm (sys)})\%$ at a nominal neutrino energy of 1.2 GeV.
Using the Rein-Sehgal model for incoherent 
production~\cite{RS2} and our present calculation of the coherent
cross section we get a coherent fraction of $\approx 5\%$
which is much below the experimental findings. It would be interesting to see
if there is a dependence of the experimental fraction on the details 
of the coherent model used in the analysis. 
It should be stressed that the theoretical prediction for coherent scattering covers only
reactions where the nucleus stays intact and does not break up during
interaction. The data on pion-Carbon scattering~\cite{Schlaile} show that nearly two-thirds of the cross section is inelastic.
Our estimate of the error in
the theoretical prediction (taking account of the model dependence of  $\sigma_{\rm incoh}$,
the extrapolation
in $Q^2$, the interpolation in fig.\ref{fig2}
and the neglect of the transverse contributions $\sigma_{R,L}$ estimated in~\cite{Paschos}) is $20\%$.

The total pion nucleus cross section scales $\sim A^{2/3}$
at least for small atomic numbers $A$~\cite{Hufner}.
If the ratio of elastic to total cross section depends only weakly on $A$ we 
then expect the elastic cross section also to scale
proportional to $A^{2/3}$.
Using the RS hadronic model the ratio of the cross sections for
Aluminum ($A=27$) and Carbon is $1.68$ for $E=2$ GeV corresponding to a 
scaling law $\sim A^{0.63}$.
The algorithm presented in the preceding section can thus probably
be simply extended to other light nuclei.
Applying the $A^{2/3}$  scaling law we obtain for Aluminum 
$\sigma_{\rm coh}^{\pi^\circ}=13.3\times 10^{-40}$ cm$^2$
at $E=2$ GeV which agrees within errors with the experimental value of
$(27\pm 17)\times 10^{-40}$ cm$^2$ measured by the Aachen Padua experiment~\cite{ACP}.

For neutrino energies $>2$ GeV the calculation using Carbon data is not applicable because
a large part of the integration then needs Carbon data for $|\bm{p}_\pi|>1$ GeV which are
not available. 
The fact that the RS prediction for pion Carbon scattering overlaps with the
empirical result in the region $|\bm{p}_\pi|> 0.7$ GeV (fig.\ref{fig2}) may be
a hint that the RS model begins to be a valid description of coherent
pion production beyond the resonance region.
Such a view would explain the impressive agreement
of the RS model with the large body of data on coherent pion production
in the energy domain $E=2..100$ GeV, as documented for example in~\cite{Charm}.



\end{document}